\documentclass[aps,prl,preprint,groupedaddress]{revtex4-1}
\usepackage[pdftex]{graphicx}
\usepackage{dcolumn}	
\usepackage{bm}			
\usepackage{here}
\usepackage[dvipdfm]{}
\usepackage{graphicx}
\usepackage{epstopdf}
\usepackage{mathrsfs}
\usepackage{amsmath}

\def\cx{Cu$_x$Bi$_2$Se$_3$}

\def\BS{Bi$_2$Se$_3$}

\def\dv{\textbf {d}-vector}
\begin{document}

\title{Manipulating the  nematic director  by magnetic fields in the  spin-triplet  superconducting state of  Cu$_{x}$Bi$_2$Se$_3$} 
\author{M. Yokoyama}
\thanks{Partly based on Master degree theses by M. Yokoyama (Feb. 2021) and H. Nishizaki (Feb. 2022), Okayama University.}
\affiliation{Department of Physics, Okayama University, Okayama 700-8530, Japan}
\author{H. Nishigaki}
\thanks{Partly based on Master degree theses by M. Yokoyama (Feb. 2021) and H. Nishizaki (Feb. 2022), Okayama University.}
\affiliation{Department of Physics, Okayama University, Okayama 700-8530, Japan}
\author{S. Ogawa}
\affiliation{Department of Physics, Okayama University, Okayama 700-8530, Japan}
\author{S. Nita}
\affiliation{Department of Physics, Okayama University, Okayama 700-8530, Japan}
\author{H. Shiokawa}
\affiliation{Department of Physics, Okayama University, Okayama 700-8530, Japan}
\author{K. Matano}
\affiliation{Department of Physics, Okayama University, Okayama 700-8530, Japan}
\author{Guo-qing Zheng}
\affiliation{Department of Physics, Okayama University, Okayama 700-8530, Japan}

\begin{abstract}
	Electronic nematicity, a consequence of  rotational symmetry breaking, is an emergent phenomenon in various new materials. 
	In order to fully  utilize the functions of these materials,  ability of  
	tuning them  through a knob, the nematic director,  is desired.
	Here we report a successful manipulation  of the nematic director, the vector order-parameter ($\textbf{{d}}$-vector),  in the  spin-triplet superconducting state of  Cu$_x$Bi$_2$Se$_3$ by  magnetic fields. 
	At  $H$ = 0.5 T, the ac susceptibility related to the upper critical field  shows a 
	two-fold symmetry in the basal plane. 
	At $H$ = 1.5  T, however, the susceptibility 
	shows a six-fold symmetry, which has never been reported before in any superconductor. These results indicate that the $\textbf{{d}}$-vector initially pinned to a certain direction is unlocked  by  
	a threshold field   to respect the trigonal crystal symmetry. %
	We further reveal that the superconducting gap in different crystals converges to $p_x$ symmetry at high fields, although it differs at low fields.
\end{abstract}
\maketitle




Skyrmion spin textures of magnets 
\cite{Pflei},  the normal state of iron-pnictides \cite{Fernandes-Shmalien-Chubukov}, and the superconducting states of 
spin-triplet  superconductors \cite{MatanoKrienerSegawaEtAl2016,Yang} and magic-angle graphene \cite{YCao}, are all nematic. Furthermore, the excitation in the vortex cores of spin-triplet superconductors can form nematic skymion-type texture \cite{Babaev}.
%
In particular, nematic  spin-triplet superconducting states  are topological \cite{FuPRL}, where
Majorana fermions (excitations) are expected to appear on edges or in the vortex cores\cite{QiZhang,Ivanov},  
which can  potentially  be applied  to  fault tolerant
non-Abelian quantum computing \cite{TopologicalQuantumComputation,quantumcomputer_KITAEV20032}.
However,  bulk spin-triplet  superconductors are still very rare. 
Carrier-doped topological insulator Cu$_{0.3}$Bi$_2$Se$_3$ \cite{MatanoKrienerSegawaEtAl2016} and ferromagnetically correlated electron system K$_2$Cr$_3$As$_3$ \cite{Yang} are recently-established spin-triplet superconductors, with the superconducting transition temperature $T_c$ as high as 6.5 K. Along with the   uranium-based candidates such as  UTe$_2$ \cite{UTe2}, they provide good platforms for the study of topological quantum phenomena. 
In order to implement these compounds in  applications, however, one still needs to better understand the physics of the spin-triplet states in these materials.

In contrast to spin-singlet state, 
 a spin-triplet superconducting state is described by the vector order parameter \textbf{{d}},  
 whose direction is perpendicular to the direction of paired  spins  and whose magnitude is the gap size \cite{Balian}. 
 In superfluid $^3$He, the \dv\ rotates freely \cite{Leggett}. In a solid, however,  the \dv\ can be  pinned to a certain direction which results in spin-rotation symmetry breaking as first found in Cu$_{0.3}$Bi$_2$Se$_3$  \cite{MatanoKrienerSegawaEtAl2016}. The pinning of the \dv\ is the origin of the observed   nematic responses \cite{MatanoKrienerSegawaEtAl2016,Yonezawa_Natphys,Sr_dope_2fold_PanNikitinAraiziEtAl2016,Asaba_Sr_PhysRevX.7.011009,Sr_dope_2fold_Du2017,FengDL,Kawai}, so that  the $\textbf{{d}}$-vector is the nematic director.
 When the   \dv\   rotates or is flipped, the magnetic response also changes and the pairing symmetry can even change, 
 giving rise to a  transition from a phase with one symmetry to another  with different symmetry.  Cu$_x$Bi$_2$Se$_3$    exemplifies such intriguing property,  where  carrier-concentration  tunes the superconducting phase from one to another with different \dv\ direction. For low doping level with $x<$0.46, the \dv\ lies in the basal plane, while for high dopings with $x\geq$0.46, the \dv\ rotates to the $c$-axis direction \cite{chiral_PhysRevB.94.180504,Kawai}, accompanying a possible nematic-to-chiral phase transition.
 Thus, a thorough understanding of the $\textbf{{d}}$-vector  and its interaction with the environment and external perturbations is important.

In this paper, we report a successful manipulation of  the $\textbf{{d}}$-vector in Cu$_x$Bi$_2$Se$_3$ by a magnetic field as small as 1 Tesla.
We synthesized  Cu-doped \BS\ single crystals with low doping rate by the electrochemical intercalating method.
Through the measurements of $^{77}$Se nuclear magnetic resonance (NMR), we confirm the small carrier concentration and less disorder/defects caused by doping of the new crystals.
We measure the ac-susceptibility in the superconducting state by rotating the sample in a magnet to change the angle between the magnetic field and the crystal $a$-axis. We further reveal the intrinsic gap symmetry of Cu$_x$Bi$_2$Se$_3$.
 
Single crystals of \cx\ were prepared by intercalating Cu into \BS\ by the electrochemical doping method described in Ref.[\onlinecite{Kawai,Kriener_PRB2011}].
First, single crystals of \BS\ were grown by melting stoichiometric mixtures of 
elemental Bi (99.9999\%) and Se (99.999\%) at 850 $^{\rm o}$C for 48 hours in sealed evacuated quartz tubes. 
After melting, the sample was slowly cooled down to 550 $^{\rm o}$C over 48 hours
and kept at the same temperature for 24 hours.
Those melt-grown \BS\ single crystals were cleaved into smaller rectangular pieces of about 14 mg.
They were wound by bare copper wire (dia. 0.05 mm), and used as a working electrode.
A Cu wire with diameter of 0.5 mm  was used both as the counter (CE) and the reference electrode (RE).
We applied a current of 10 $\mu$A in a saturated solution of CuI powder (99.99\%) in acetonitrile (CH$_3$CN).
The obtained crystals samples were then annealed at 560 $^{\rm o}$C for 1 hour in sealed evacuated quartz tubes, and quenched into water.
After quenching, the samples were covered with epoxy (STYCAST 1266) to  avoid deterioration.
We have confirmed that the epoxy does not have extrinsic effect on the physical properties such as $T_{\rm c}$ or upper critical field $H_{\rm c2}$. 
The Cu concentration $x$ was determined from the mass increment of the samples.
To check the superconducting properties, dc susceptibility measurements were performed using a superconducting quantum interference device (SQUID) with
the vibrating sample magnetometer (VSM).
%
The $^{77}$Se-NMR spectra were obtained by the fast Fourier
transformation of the spin-echo  at a field of $H_0 $ = 1.5 T. 
The Knight shift $K$ was calculated using nuclear gyromagnetic ratio $\gamma_{\rm N}$ = 8.118 MHz/T for $^{77}$Se.
%
The
ac susceptibility  was measured by the inductance of an in-situ NMR coil. 
Angle-dependent measurements 
were performed by using a piezo-driven rotator (Attocube  ANR51) equipped with Hall  sensors  to determine the angle between  magnetic field and  crystal axis. We estimate that the error  in the angle determination is less than 1 degree.

\begin{figure}[htbp]
	\includegraphics[clip,width=80mm]{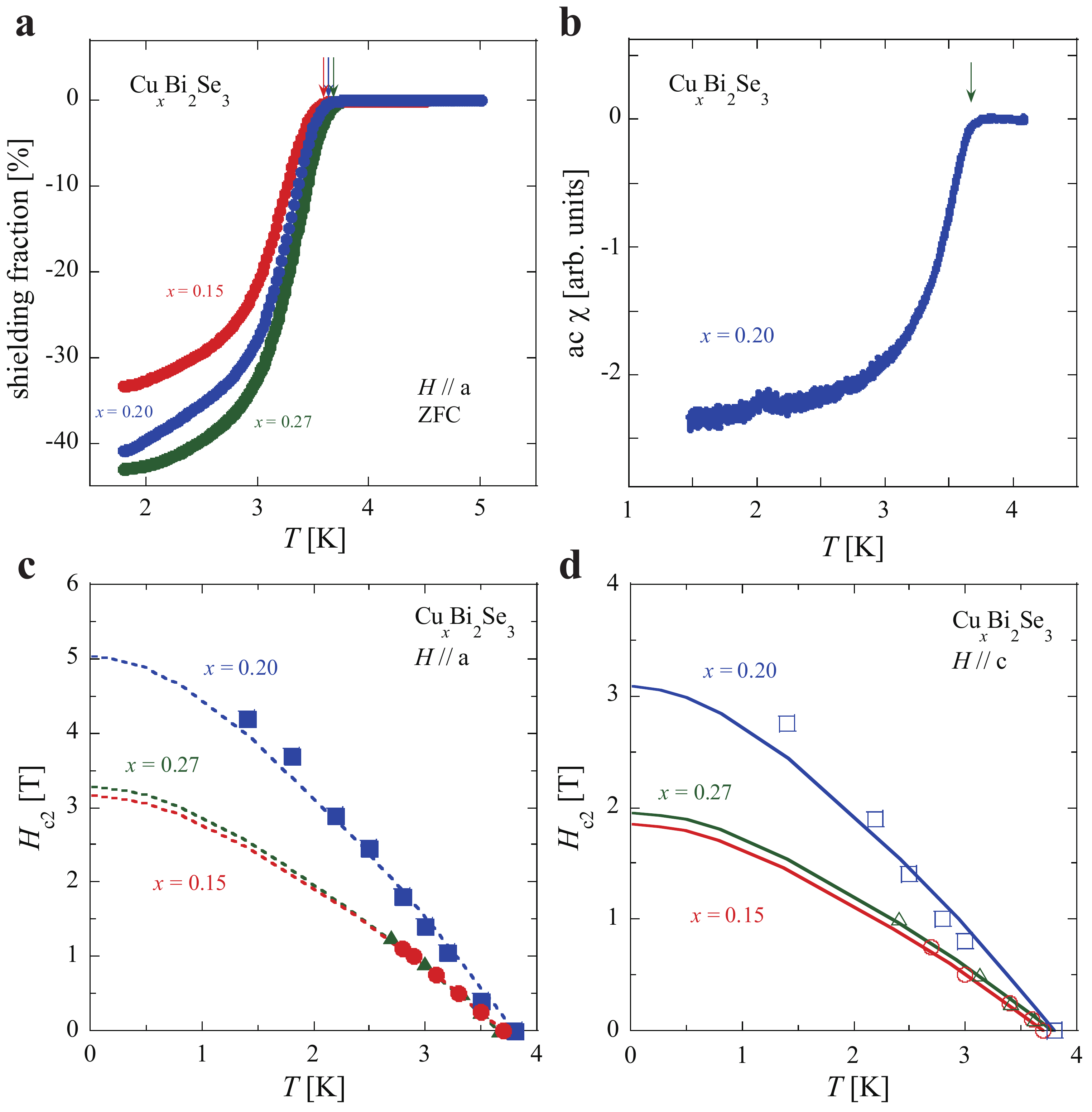}
	\caption{
		(color online)(a) Superconducting transition and the shielding fraction of Cu$_{x}$Bi$_2$Se$_3$ ($x$=0.15, 0.20 and 0.27). Arrows indicate $T_c$ for the samples. (b) ac susceptibility for the sample with $x$=0.20. A small hump around $T$=2.1 K is due to the environmental change associated with the superfluid transition of liquid Helium. 
		(c) upper critical field $H_{\rm c2}$ for $H\parallel a$ for the three samples, (d)  $H_{\rm c2}$ for $H\parallel c$ for the three samples.
	}
\end{figure}
%
%
Figure 1(a) shows the superconducting transition  of the three samples Cu$_{x}$Bi$_2$Se$_3$ ($x$=0.15, 0.20 and 0.27) obtained by  dc susceptibility measurements. 
The $T_c$ is 3.6, 3.8 and 3.6 K for $x$=0.15, 0.20 and 0.27, respectively, which is 
 higher than  $T_c$ of other $x$ concentrations reported for $x\geq$0.28, following a general trend that $T_c$ increases with decreasing $x$ \cite{Kriener_PRB2012,Kawai}.
The shielding fraction at $T$=1.8 K for $x$=0.20 and 0.27 exceeds 40\% which is among the highest value reported so far. 
In our case, demagnetization is negligible as the magnetic field is applied parallel to the plate.
Figure 1(b) shows the data  for $x$=0.20 as a representative example for  ac susceptibility. 
Figure 1(c) and Fig.1(d) show the upper critical field $H_{\rm c2}$ for $H\parallel a$ and $H\parallel c$, respectively.  The $H_{\rm c2}$ for $x$=0.15 and 0.27 is comparable to that reported previously by Kriener et al \cite{Kriener_PRB2012}, but $H_{\rm c2}$ for $x$=0.20 is much higher.
\begin{figure}[htbp]
	\includegraphics[clip,width=70mm]{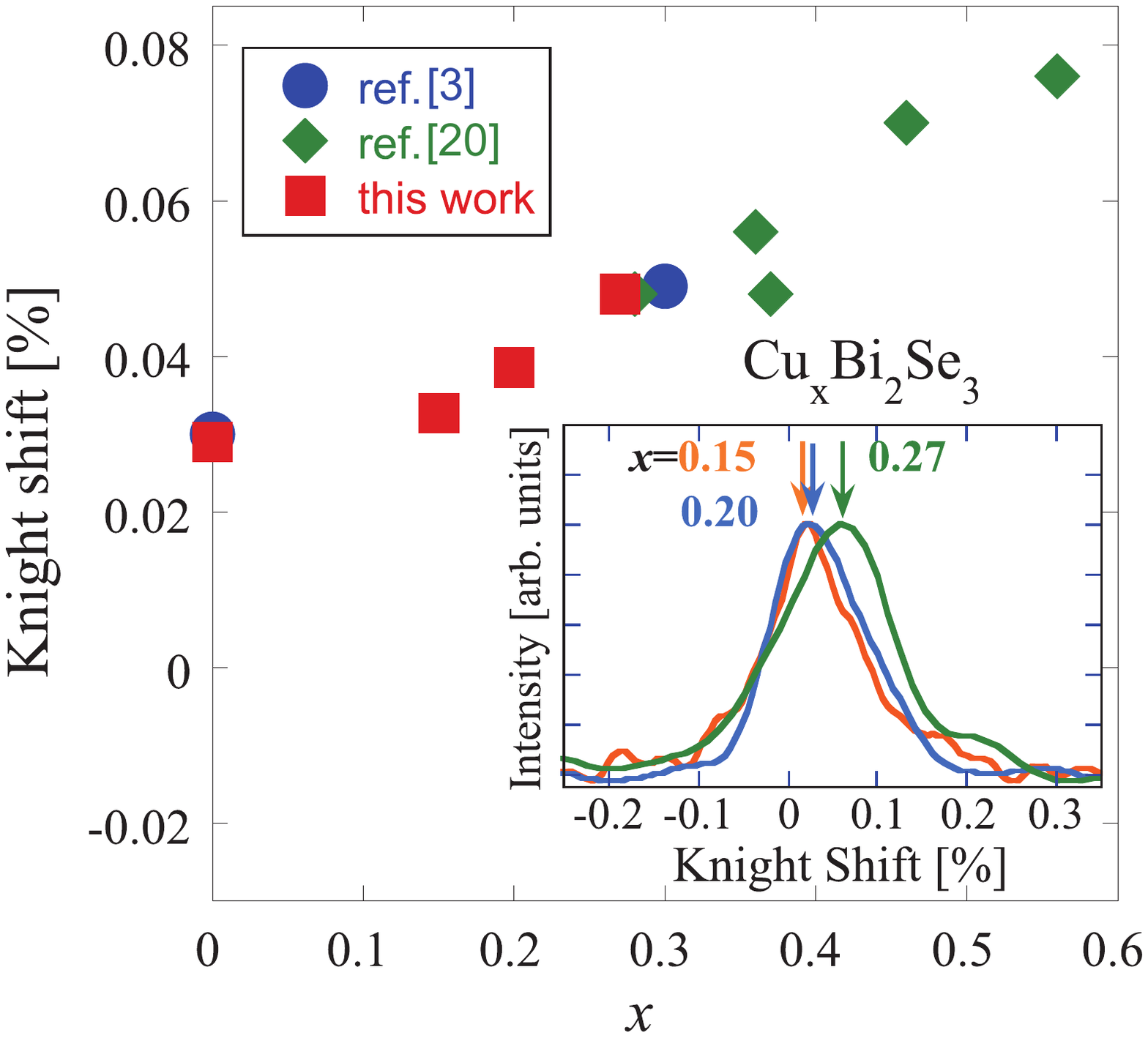}
	\caption{\label{spec}
		(color online)	The $^{77}$Se Knight shift as a function of nominal Cu-content $x$. The inset shows the NMR spectrum  for $x$=0.15, 0.20 and  $x$=0.28. The arrow indicates the gravity center of the spectrum from which the Knight shift was extracted.
	}
\end{figure}
In Fig. \ref{spec}, we show the $^{77}$Se Knight shift as a function of Cu-content $x$ for these samples, together with the data for other samples reported previously \cite{MatanoKrienerSegawaEtAl2016,Kawai}. The Knight shift, which is proportional to the density of states, decreases with decreasing $x$, 
being consistent with a smaller carrier concentration for the new crystals  than that for the previous ones with $x\geq$0.28.
 The $^{77}$Se-NMR spectrum taken at $H$=1.5 T and $T$=3.0 K above $T_{c}(H)$ was shown in the inset to the figure. 
The spectrum  becomes sharper  with decreasing $x$, with the full width at half maximum (FWHM) of 14.4 kHz and 14.7 kHz for $x$=0.15 and $x$=0.20, respectively,   which is narrower than that for  $x$=0.27 (16.5 kHz) and $x\geq$0.28 \cite{MatanoKrienerSegawaEtAl2016,Kawai} and thus ensures that these low-doping samples have a less disorder.   
%

\begin{figure}[htbp]
	\includegraphics[clip,width=90mm]{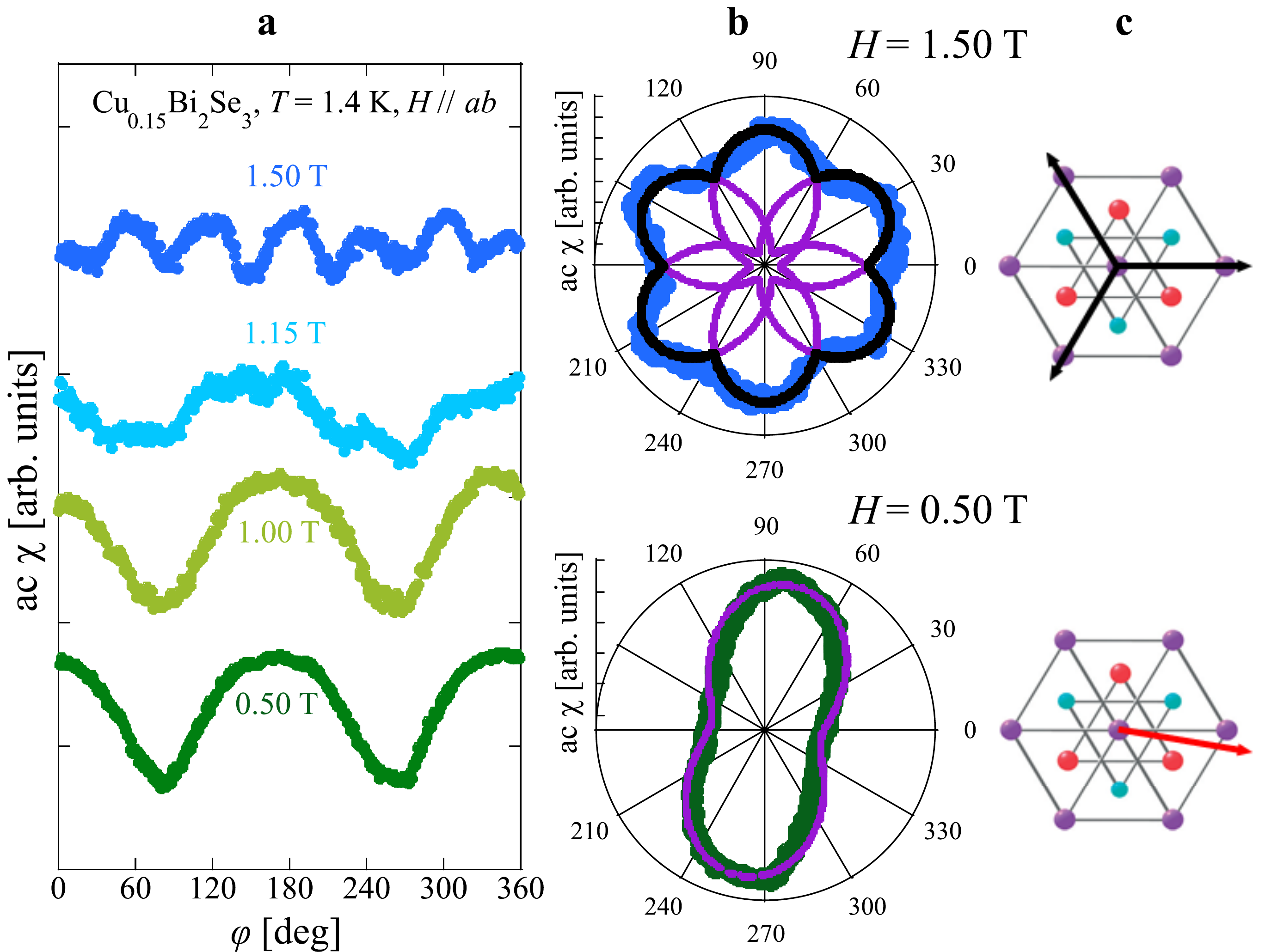}
	\caption{\label{x=0.15}(color online) 
			Field- and angle-evolution of the diamagnetism and $\textbf{{d}}$-vector revealed by the ac susceptibility measurement for $x$=0.15. (a) In-plane angle dependence of the diamagnetism  at $T$=1.4 K measured by the ac susceptibility under various fields.	(b) Polar plot of the ac susceptibility at $H$=0.5 T and 1.5 T, respectively. The purple curve in the lower panel is a simulation of an anisotropic $H_{\rm c2}$ formula (see text). In the upper panel, the black curve connects the outer envelop of the three ellipses.  (c) Illustration of the  $\textbf{{d}}$-vector(s) by arrow(s) at different fields. The purple balls depict Bi, and the red and green balls represent Se in the bird-viewed basal planes.
	}
\end{figure}

\begin{figure}[htbp]
	\includegraphics[clip,width=90mm]{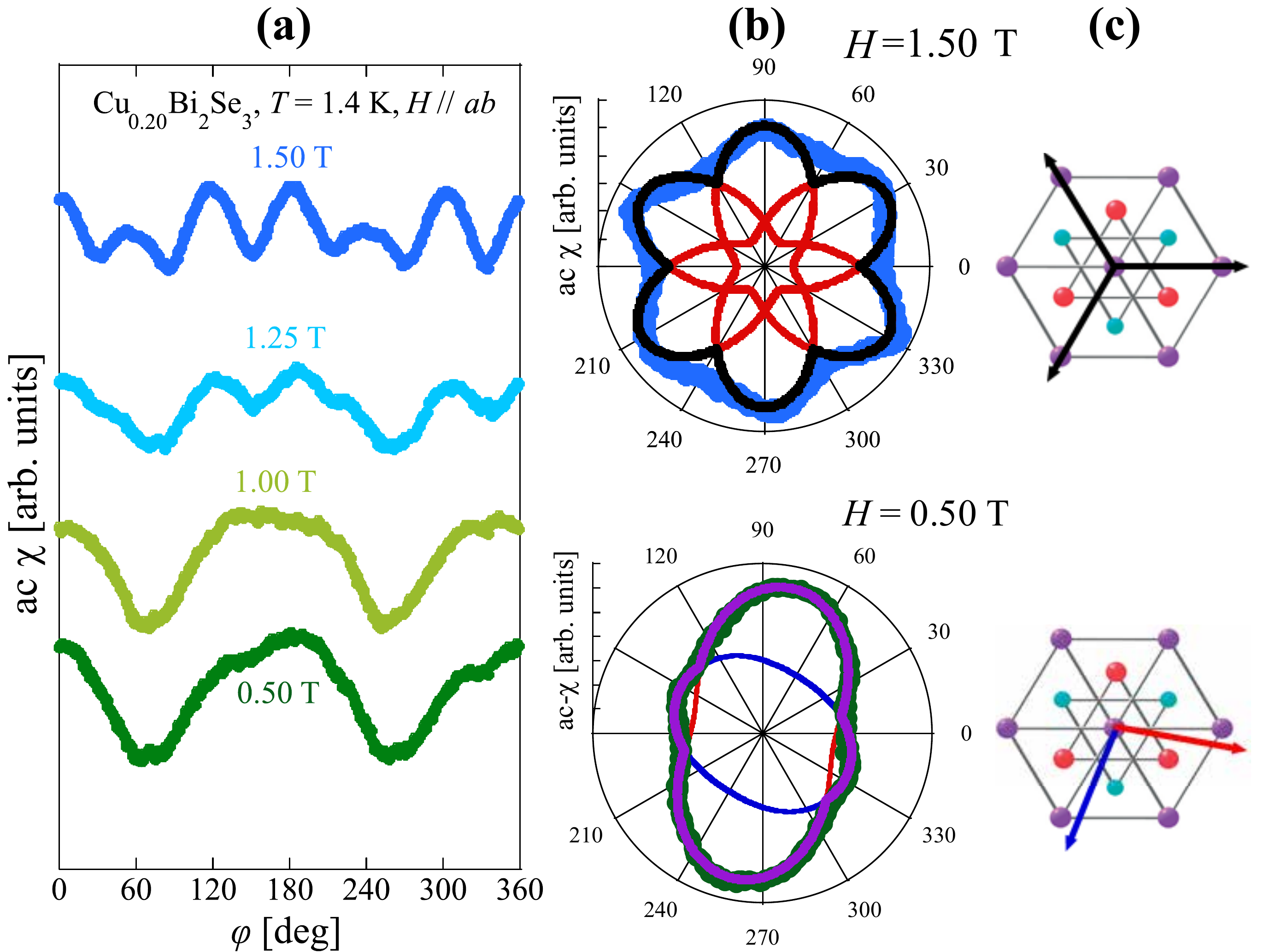}
	\caption{\label{x=0.2}(color online) 
		(a) In-plane angle dependence of the ac susceptibility of the $x$=0.20 crystal at $T$=1.4 K for various fields.	(b) Polar plot of the ac susceptibility at $H$=0.5 T and 1.5 T. In the lower panel, red and blue curves are the simulations of the ac $\chi$ from two different domains. The thick purple curve is the outer envelop of the two ellipses. For other versions of the simulation curves, see Supplemental Material \cite{Supp} (c) Illustration of the  $\textbf{{d}}$-vector direction(s).
	}
\end{figure}

\begin{figure}[htbp]
	\includegraphics[clip,width=90mm]{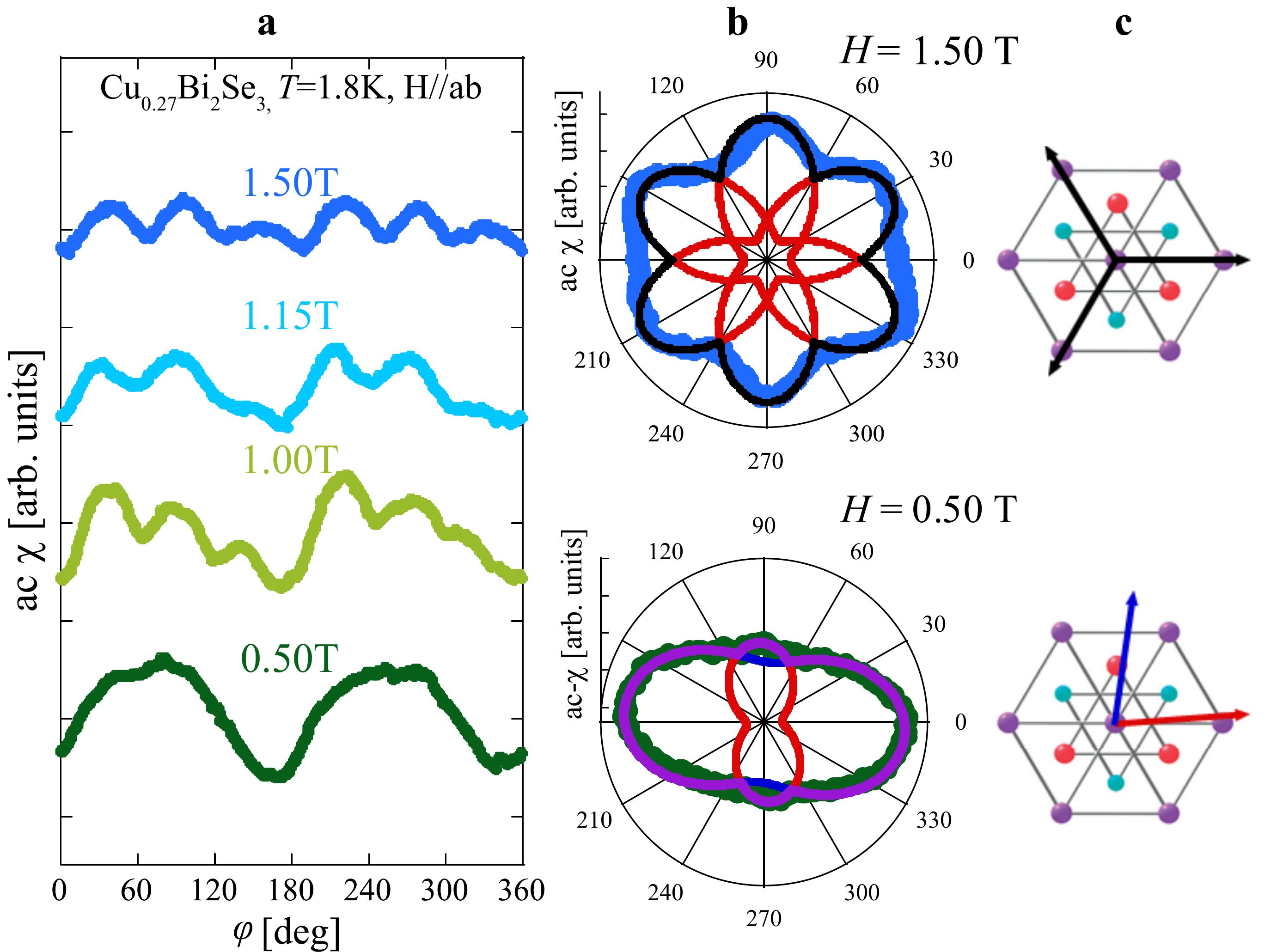}
	\caption{\label{x=0.27}
		(color online)	Field- and angle-evolution of the diamagnetism and $\textbf{{d}}$-vector  for $x$=0.27.
		 (a) In-plane angle dependence of the diamagnetism  at $T$=1.8 K measured by the ac susceptibility under various fields for $x$=0.27.	(b) Polar plot of the ac susceptibility at $H$=0.5 T and 1.5 T, respectively. The captions for the curves are the same as Fig. 4(b). (c) Illustration of the  $\textbf{{d}}$-vector direction(s) at different fields.
	}
\end{figure}

Figures   \ref{x=0.15}-\ref{x=0.27} 
show the main results of this work.
Figures   \ref{x=0.15}(a),  Fig.\ref{x=0.2}(a) and Fig.\ref{x=0.27}(a) depict the angle $\varphi$ dependence of the diamagnetism measured by ac susceptibility at $T$=1.4 K (see Fig.1(b) for an example) under various magnetic fields. 
Here $\varphi$ is the angle between the crystal $a$-axis and the magnetic field. The pre-determined $a$-axis direction is set to be $\varphi$=0 degree.
The plotted quantity is related to  $H_{\rm c2}$; the larger  $H_{\rm c2}$, the larger diamagnetism at a fixed temperature and field.
At $H$= 0.5 T, a (nearly) two-fold symmetry is observed, in agreement with previous Knight shift \cite{MatanoKrienerSegawaEtAl2016} and $H_{\rm c2}$ \cite{MatanoKrienerSegawaEtAl2016,Yonezawa_Natphys,Kawai} measurements. 
At $H$=1.15 T (1.25 T for $x$=0.20 and 1 T for $x$=0.27), new components emerge in the oscillation, and surprisingly, there emerges six minima in the oscillation   at $H$=1.5 T. The situation is better visualized in the polar plots in Fig.   \ref{x=0.15}(b), Fig.   \ref{x=0.2}(b) and Fig. \ref{x=0.27}(b). A  six-fold symmetry is clearly seen at  $H$=1.5 T.

The data in the lower panel of  Fig.\ref{x=0.15}(b) can be fitted by a phenomenological formula, 
	-$\chi$ = - $\frac{|\chi_{max}|}{\sqrt{cos^2(\varphi-\theta)+c\cdot sin^2(\varphi-\theta)}}$.
Here, $|\chi_{max}|$ is the largest diamagnetic susceptibility and  $\theta$ is the angle between the elliptical direction and the $a$-axis, and $c$ is a number parameter that determines the shape of the  ellipse. The parameter $c$ obtained from the fittings in  Fig.\ref{x=0.15}(b), Fig.\ref{x=0.2}(b) and Fig.\ref{x=0.27}(b) is 1.3, 1.2 and 1.4, respectively. A smaller $c$ makes the shape more peanuts-shape like while a larger $c$ makes the shape more ellipse-like. This formula was originally developed  to describe the $H_{\rm c2}$ anisotropy \cite{Hc2-anisotropy} and was also adapted in Ref. \cite{Sr_dope_2fold_PanNikitinAraiziEtAl2016}.

Although the data for $x$=0.15 
agree well with the simulation of a single ellipse (a single component), the data for $x$=0.20 and 0.27 do not. There is a small second component, as can be seen in the lower panels of  Fig.\ref{x=0.2}(b) and Fig.\ref{x=0.27}(b). For $x$=0.20, the $\theta$ for the small, second component is  340$^{\circ}$, while it is 80$^{\circ}$ for the main component. For $x$=0.27,  the  $\theta$ is  355$^{\circ}$  for the main component, while it is 95$^{\circ}$ for the second component.
In either case, the sub-dominant component disappears at high fields beyond 1.5 T,  which indicates that it arises from a different crystal domain with a lower 
$H_{\rm c2}$  (see below for more discussion). 
The single domain seen in $x$=0.15 can be understood as due to its lower doping which makes the crystal more homogeneous as evidenced by the sharper NMR spectrum.
Theoretically, two degenerate gap states corresponding to $p_x$ and $p_y$ were proposed by Fu \cite{Fu_CuxBi2Se3_PhysRevB.90.100509}.
The main component for $x$=0.15 and $x$=0.20 is compatible with  $p_x$, but that for $x$=0.27 is compatible with  $p_y$ symmetry. In the previous work, $H_{\rm c2}$ measurements found that  $x$=0.3 and 0.37 correspond to $p_x$ while $x$=0.28 and 0.36 correspond to $p_y$ \cite{MatanoKrienerSegawaEtAl2016,Kawai}. 

Looking into the detail of the lower panels of Fig. \ref{x=0.15}(b), Fig. \ref{x=0.2}(b) and Fig. \ref{x=0.27}(b), one notices that the elliptical axis is tilted away from the high-symmetry (crystal axis) direction. As the  $\textbf{{d}}$-vector is perpendicular to the main axis of the  $H_{\rm c2}$ ellipse 
\cite{MatanoKrienerSegawaEtAl2016,Kawai},  we  illustrate the corresponding $\textbf{{d}}$-vector for the present three crystals in the lower panels of Fig. 3(c), Fig. 4(c) and Fig. 5(c). For $x$=0.20, the $\textbf{{d}}$-vector for the main component is tilted from the high symmetry axis by $\delta$=-10$^{\circ}$, while the tilting angle is  $\delta$=10$^{\circ}$ for the sub-dominant component. 
So the two components can be regarded as belonging to the same gap symmetry in view of the trigonal crystal symmetry.
For $x$=0.27, the $\textbf{{d}}$-vectors is  85$^{\circ}$, which is close to 90$^{\circ}$, for the main component, while it is  5$^{\circ}$ for the minor component. Therefore,  the two components correspond to different symmetries that are orthogonal to each other.
%
%
%
We believe that the   cause for the tilting 
from the high-symmetry axis direction is  phonon-mediated interaction. 
Hecker and  Fernandes \cite{phonon} recently proposed that the competition between a quadratic phonon-mediated interaction $E_{\rm nem-ph}$ and the cubic
nematic anharmonicity can make the nematic director 
deviate from the high-symmetry direction. The intrinsic nematic anharmonic cubic interaction $E_{\rm nem}$ favors the nematic director to align parallel to the high-symmetry directions, while the phonon-mediated non-analytic quadratic interaction 
prefers the nematic director to
align to the directions farthest away from
the high-symmetry directions.

Hecker and  Fernandes further 
pointed out that the phonon interaction and thus the nematic director ($\textbf{{d}}$-vector) rotation will have an impact on domain formations \cite{phonon}. If the $\textbf{{d}}$-vector is along the high symmetry axis, the three equivalent directions have the same angular separation,
and  the surface energies between any two domains
are equal. As a result, when one direction among the three is
chosen as the majority-domain $\textbf{{d}}$-vector, the other two directions will be
randomly picked as minority-domains $\textbf{{d}}$-vector \cite{phonon}. However, 
if the $\textbf{{d}}$-vector is rotated by an angle $\delta$ away from the crystal axis, in order to minimize the surface energy, then there will only be one type of minor domain 
with the   $\textbf{{d}}$-vector rotated by an angle of -$\delta$. This is exactly we have found for the $x$=0.20  crystal.







At low fields, the two-fold nematic behavior in the physical quantities is well understood as due to the \dv\ pinning to a certain direction by, {\it e.g.}, spin-orbit coupling promoted by disorder or defects,
although there are three equivalent directions favored by the $\textbf{{d}}$-vector.    
 In superfluid $^3$He, the $\textbf{{d}}$-vector rotates freely as there is no lattice.
As can be seen in Figs.3-5, the  oscillation in $H_{\rm c2}$ restores a  six-fold symmetry at high fields of $H\geq$1.5 T, compatible with 
the trigonal crystal-lattice  symmetry. Namely, the three $a$-axis directions become equivalent at high fields, as 
 depicted in the upper panels of Fig. 3(c), Fig.4(c) and Fig. 5(c).
 This means that the  $\textbf{{d}}$-vector is depinned by the application of a large magnetic field. 
 We should note that a possible change in vortex lattice structure cannot explain the observed symmetry change. Firstly, any vortex lattice cannot give rise to a two-fold symmetry  in physical quantities. Secondly, a higher order effect could lead to a response of six-fold symmetry at high fields in principle, but in such case a circular-like shape should appear at low fields \cite{Ichioka},  which is not seen in our experiments.  
 
 The simplest interpretation of the symmetry transition in  $H_{\rm c2}$ is through the total energy $E_{\rm tot} = E_{\rm nem} + E_{\rm nem-ph}+ E_{\rm Zeeman}$ including Zeeman energy $E_{\rm Zeeman}$. Above a threshold value $H_{\rm pin}$, the $\textbf{{d}}$-vector 
  traces the rotating magnetic field as to gain Zeeman energy which is the largest when the $\textbf{{d}}$-vector is perpendicular to the field. 
 The energy gain by $E_{\rm nem}$ is the largest when the $\textbf{{d}}$-vector is along the three high-symmetry directions.
  The $H_{\rm pin}$ is 1.0$\sim$1.2 T in the case of Cu$_{x}$Bi$_2$Se$_3$. This is the first case, to our knowledge, that the $\textbf{{d}}$-vector can be manipulated by the magnetic field.
  Also, a six-fold symmetry of  $H_{\rm c2}$ or its related physical quantities has never been observed before in  any superconductor, although theories have pointed out that a trigonal superconductor with a two-component order-parameter may show such property under certain conditions \cite{Fu-Hc2,Mineev}. Therefore, our result is supplemental to the Knight shift measurement which revealed a  spin-triplet state in Cu$_{x}$Bi$_2$Se$_3$.

The most intriguing feature is that,  although the long axis direction  of the ellipse for the angle-dependent Meissner diamagnetism is tilted way from the high symmetry axis directions  of the  trigonal crystal lattice at low fields, 
with $\theta$=80$^{\circ}$ for $x$=0.15 and  $x$=0.20, 
 it is restored to the high symmetry axis directions at high fields $H\geq$1.5 T. 
 This means that the phonon-induced interaction  $E_{\rm rem-ph}$ is also smaller than the Zeeman interaction. 
Such symmetry at high magnetic fields is compatible with the so-called $\Delta_{4x}$ ($p_x$) state proposed \cite{Fu_CuxBi2Se3_PhysRevB.90.100509}. 
In contrast, the ellipse for $x$=0.27 at low fields has an elliptical direction along  $\theta$=355$^{\circ}$ and thus the gap
 is compatible with the other state, the so-called $\Delta_{4y}$  ($p_y$) state. 
The previously reported crystals of $x$=0.28 and 0.36 also showed such symmetry.
Most strikingly and surprisingly,  the $\Delta_{4y}$  ($p_y$)  compatible elliptical shape at low fields 
is also restored to the symmetry of  $\Delta_{4x}$ ($p_x$) at high fields, as seen in the upper panels of Fig.5(b) and Fig.5(c).
We have investigated more than 16 crystals with 0.15$\leq x \leq$0.40, and found that all of them become  compatible with the $\Delta_{4x}$ state after the \dv\ is depinned at high magnetic fields, although  the  $\theta$ is different from crystal to crystal at low fields and some of them  are consistent with the $\Delta_{4y}$ symmetry.
This implies that the intrinsic gap symmetry is $\Delta_{4x}$. We speculate that the  $\Delta_{4y}$ appearing  in some cases  is accidental due to local defects caused by the quenching process. This is an open question that needs to be addressed in the future. 

In conclusion, for the first time,  we have successfully manipulated the nematic director by magnetic fields.  
The diamagnetism in single crystal samples of  \cx\ measured by  ac susceptibility shows a two-fold symmetry with respect to the angle between the field and the crystal $a$ axis at a low field $H$ = 0.5 T, with the direction of largest diamagnetism slight deviated from the crystal axis. 
At high fields of
 $H \geq$ 1.5 T, however, the ac susceptibility shows a six-fold symmetry, exactly matching the   crystal axes. These results indicate that the $\textbf{{d}}$-vector initially pinned to a certain direction is unlocked  by  the magnetic fields above a threshold value to trace the field. The six-fold symmetry in  $H_{\rm c2}$ or its related physical quantities  was expected for a spin-triplet superconductor with two-component order parameter, but has never been observed before. Thus, our results are supplemental to the Knight shift result which found  that Cu$_x$Bi$_2$Se$_3$ is a spin-triplet superconductor.
 Our work further reveals the $p_x$ gap  symmetry at high fields for all samples, irrespective of different symmetries at low fields, indicating that this is the intrinsic gap symmetry of Cu$_x$Bi$_2$Se$_3$.

\section{Acknowledgments}
We thank Y. Inada for help in Laue diffraction measurements, S. Kambe  for advice in crystal growing,  S. Kawasaki for help in susceptibility measurements, and M. Ichioka and R. Fernandes   for useful discussions. 
This work was supported in part by the JSPS Grants No. 19H00657, No. 20K03862 and No. 22H0448 (Grant-in-Aid for Scientific Research on Innovative Areas “Quantum Liquid Crystals”).
%


%


\end{document}